\documentclass[preprint]{revtex4}

\begin{document}

\title{Chaos and complexity in astrophysics}
\author{Oded Regev$^{1,2}$}
\affiliation{$^{1}$Department of Physics, Technion-Israel
Institute of Technology, Haifa, Israel \\
 $^{2}$ Department of Astronomy, Columbia University, New York,
NY 10027, USA}
\begin{abstract}
\noindent Methods and techniques of the theory of nonlinear dynamical systems and
patterns can be useful in astrophysical applications. Some works on the subjects
of dynamical astronomy, stellar pulsation and variability, as
well as spatial
complexity in extended systems, in which such approaches have already been utilized,
are reviewed. Prospects for future directions in applications of this kind
are outlined.\\
\vskip2.cm
\centerline{\em to appear in the}
\centerline{\bf Springer Encyclopedia of Complexity and System Science}
\centerline{\tt http://refworks.springer.com/complexity}
\end{abstract}
\maketitle
\section{Apologies and Explanations}

I apologize for not being able to provide the readers with the
pleasure of downloading this preprint directly from the {\tt arXiv:astro-ph} archive.
Intensive efforts on my part to post this preprint on the
{\tt arXiv:astro-ph} have unfortunately failed. It is undoubtedly due to my ignorance
in the advanced methods of computer science, required for this task. I have posted papers
before on the preprint archive, but this time my .eps figure files were either "too large"
or "not sufficiently portable" so that the posting was repeatedly rejected despite my
heroic efforts to improve the "encoding", "compression" etc. of the miserable
.eps files.

Neither Cambridge University Press (CUP) nor Springer-Verlag (SV) had any problems with
{\em the same figure files}. They were happily
accepted and were either already published (in my CUP book) or will be published in the future
(in the SV Encyclopedia). My old laptop also easily compiled the LaTeX file, embedded these
figures and produced a .pdf file of $\sim 2.1$ MB (obviously prohibitively large for a mere
30 page article, with 11 figures, and reflecting the author's incompressibility).

Interested readers are invited to download the manuscript in .pfd (with the figures!)
from my home-page, residing on the disk
of a very modest computer at my Department:
{\tt http://physics.technion.ac.il/$\sim$regev/ccastro.pdf}\footnote{
The links in this document are non-clickable. Please re-type them in your browser. Sorry, I
can't figure out how to type the "tilde" character as text in TeX.}
I hope that a direct download of a .pdf file, without it being created first by the sophisticated
software and hardware of the {\tt arXiv}, will be sufficiently adequate for most purposes.

\section{Acknowledgement}

I am grateful to all those colleagues of mine (too numerous to be all mentioned here, lest the file
be too large) who tried, unsuccessfully, to help me in this endeavor.

This work was not supported by any grant and mere words of encouragement to {\tt regev@physics.technion.ac.il}
will be highly appreciated.

\end{document}